# A DSS framework for maintaining relevant features of Small Business B2C Websites

**Madhury Khatun**
College of Business
Victoria University
Melbourne, Victoria
Australia
Email: madhury.khatun@live.vu.edu.au

**Shah Jahan Miah**
College of Business
Victoria University
Melbourne, Victoria
Australia
Email: shah.miah@vu.edu.au

## Abstract

Managers are heavily engaged in strategic decision-making for businesses, particularly in a changing environment. One of the most important decisions for online small businesses, as part of their strategic planning, is selecting relevant features on their websites, both to attract and interact with consumers. However, only a few Australian small businesses use strategic tools for selecting their website features. As a result, businesses lose potential domestic sales in the business-to-consumer (B2C) sector. The aim of this study is to determine the relationship between factors that influence consumers' online purchasing, and owner/manager strategic decisions in selecting relevant features for websites. Results from employing qualitative case studies with small business owner/managers, and a content analysis of website features, inform the design of a Decision Support Systems (DSS) framework. This may assist owner/managers' strategic decisions to implement competitive features on B2C websites that ultimately attract more consumers.

## 1. INTRODUCTION

Managerial characteristics are heavily engaged in strategic decision-making, to address changing demands in the business environment (Baizyldayeva et al. 2013; Eisenhardt 1989; Miah et al. 2014). Managers as decision-makers analyse problems and create opportunities within the business environment (Hall 2008). IT (information technology) artefact design research for managerial decision-support solutions is therefore acknowledged to improve business strategies (Hevner et al. 2004; Miah et al. 2014). Business strategy has become an important area to continuously meet the demand of the change in external environment and organisations address such needs internally for maintaining their business performance (St-Jean et al. 2008). Bergeron et al. (2004) found that small business strategies without IT alignment provide a lower level of business performance. It is certain because, in most of the cases, small business owner/managers are poor planners and their goals are unclear, inadequately defined and short-ranged rational (Mazzarol 2004). Only a few small-businesses use IT artefact as strategic decision support tools for improving business decision making (Duan and Xu 2005; Sexton and Van Auken 1985).

In the online small-business environment, millions of consumers interact directly with companies on the Web and evaluate products and services on many websites until choose one sites (Haag and Cummings 2009). Therefore, an effective support of strategic decision is of paramount importance for small-businesses to addressing the changing environment and attracting consumers. However, in Australia, many consumers purchase from overseas websites that creates the potential loss of online sales of small businesses in the business –to- consumer (B2C) sector. What are the factors that bring consumers in overseas websites? What issues exist in Australian websites and what could be a solution to attract consumers? The objectives of our study are to find answers to the questions and provide a potential research solution through designing an IT artefact for this growing need. To achieve such objectives, it is important to analyse the internal and external business environmental factors for business decision-making (Mintzberg et al. 1976). These factors, in particular the external business environmental factors, such as customer demands, general economic condition, regulations, new technology and competitions, are critical to maintain satisfactory business performance (Beynon-Davies 2013; Pedersen and Sudzina 2012). It is a critical aspect for small businesses to address the external business environment as they have the limited support and capability.

Although many owner/managers of online small businesses have developed business strategies, they are irrelevant in most cases, and limited in their web presence (Fisher et al. 2007). The lack of strategy relates particularly to the setting-up features on their websites (Fisher et al. 2007). Small businesses often find it difficult to implement technologies due to resource constraints (Raymond 1985). Typically, owner/managers are not sufficiently knowledgeable to adopt advanced technologies as a strategic decision-making support tool.

Few studies in the DSS research domain have focused on technological improvements, particularly for integrating the online business process in a B2C e-commerce environment (Al-Qaed 2008; Jiyong and Pu 2006). There is a little research yet, to assist the small business owner/manager in making strategic decisions, especially in identifying competitive features on their business websites. For example, Al-Qaed (2008) employed the DSS approach in the B2C e-commerce environment; however, the study suffers from lack of support for options for strategic decision-making in a small business context. Therefore, in our paper, we outline requirements for designing a DSS solution framework (as an IT solution artefact) that could offer potential applications for small business owner/managers, especially in making strategic decisions to select competitive features on their Websites to achieve a better business outcome. DSS research and practice are significant in operating effective businesses, as they facilitate options for human decision-making processes into the organisational context (Angehrn and Jelassi 1994).

Over past decades, many DSS applications have been introduced to support managers in business decision-making processes (Arnott and Pervan 2008; Clark et al. 2007; Miah et al. 2014). At the enterprise level, common applications of a DSS tool can be viewed for a cash-flow analysis, improving product performance, and analysing resource allocation needs (Magee 2007). However, the extent of DSS use is generally limited and varies considerably among firms in the small-to-medium enterprise (SME) sector; particularly DSS use for strategic decisions in small businesses, which still represents both a challenge and an opportunity for DSS professionals and researchers (Duan and Xu 2005). The application of DSS also provides potential benefits to small businesses (Duan and Xu 2005; Power 2010). The most common DSS applications for small organisations are report and query applications, forecasting tools, analytical computer models and frequent buyer applications (Power 2010). Small business owner/managers were found to be more successful when developing their own DSS applications (Raymond and Bergeron 1992).

In addition, for management decision-making concepts, many decision-making models have been developed for organisational decision-making processes at various phases. For example, Simon (1960) proposed a decision model that describes the decision-making process in three main phases. These phases are Intelligence, Design and Choice. In the intelligence phase, decision makers recognise the problem, need or opportunity, data collection, analysis and exploration. This is also when decision makers collect information and knowledge from internal and external sources, and evaluate the knowledge for their organisational purposes (Holsapple 2008). In the design phase, decision makers design model(s) and search for alternatives. In the choice phase, they evaluate the alternatives generated in the previous phase. However, Mintzberg et al.'s (1976) decision-making model was considered the most useful model for investigating managers' strategic decision processes (Alalwan and Weistroffer 2011; 2014). This model empirically supports dealing with decision-making problems, and opportunities for developing decision-support aids (Alalwan and Weistroffer 2011; Huber 2013). This model also analyses internal and external business environmental factors for strategic business decision-making purposes (Mintzberg et al. 1976). Therefore, many researchers (Alalwan and Weistroffer 2011; Huber 2013; Kowalczyk and Buxmann 2014; Xueli and Wang 2012) have employed this model in the past for improving managerial business decision-making. We consider this model as small business owner/managers are the principal decision makers (Parker and Castleman 2007; 2009). This following paragraph is also done, in the list.

Alalwan and Weistroffer (2011) employed this model for the enterprise content management including the web content management, the electronic record management, the workflow management and the document management process. Huber's (2013) study used this model for computer-based decision-aiding technologies for the organizational design, problem, and opportunity identification, and the decision-making process. Xueli and Wang (2012) conducted a study in Australia that employed this model for strategic decision-making of SMEs in the manufacturing sector. The objectives of our study are to identify the factors that influence consumers in making purchase from overseas websites and apply the understanding to a potential design solution. In our paper, we propose such solution through designing a conceptual DSS solution framework that is based on Mintzberg et al.'s (1976) model for strategic decision making in the small business B2C e-commerce environment. We extend the existing model by adapting Miah et al.'s (2014) user centred DSS model for developing a final DSS solution framework. For designing DSS, this study will use Miah et al.'s (2014) model by employing design science research (DSR) method. For problem identification, we will use case study and web content analysis of small business in B2C sector in Australia. It is anticipated that the proposed solution has potential to assist decision making related to the selection of relevant competetive web features.

The paper is organised as follows. The next section describes the research background. The following two sections then introduce the theoretical framework and the research design we adopted, and the outcomes. The section after that includes data analysis and partial findings, followed by a conceptual DSS solution framework, discussions and conclusions.

## 2. BACKGROUND: RESEARCH PROBLEM

The demands of B2C e-commerce continue to grow, as more people shop online (Poon and Lau 2006). B2C e-commerce is one of the main forms of e-commerce, where online transactions are made between businesses and individual consumers (Al-Quad 2008). In the online business environment, millions of consumers interact directly with companies on the Web and evaluate products and services on many websites until they choose one site (Haag and Cummings 2009). Therefore, an effective support for strategic decision-making is of paramount importance for addressing the changing environment and at the same time attracting consumers. E-commerce is considered to provide substantial benefits to a business, and the most important benefit is an online environment for buyers and sellers. It also provides information and communication channels for buyers and sellers to complete the buying and selling process, and information and services over the Internet. In the B2C e-commerce environment, buyers have benefits with convenience access to seller websites, and sellers have benefits in selling their products anytime and anywhere (Al-Qaed 2008). It also provides many benefits to small and medium-sized enterprises (SMEs) as a relatively efficient and effective channel for information provision and exchange, advertising, marketing and completing transactions. Also, in some cases, it facilitates the distribution of goods and services to trading partners and customers (Marshall and McKay 2002). These benefits create close relationships between companies and customers (Dubelaar et al. 2005).

Although e-commerce has been providing substantial benefits to businesses, including small businesses, the majority of small businesses in the B2C sector in Australia have not yet received these benefits, as many Australian consumers buy products from overseas websites instead of local websites (Todd 2012; Deloitte 2013; Frost and Sullivan 2012; Irvine et al. 2011; PwC 2012; Sensis 2013; Sivasailam 2012), resulting in significant loss of potential on-line domestic sales in Australia (Todd 2012; Ewing 2011; Wright 2012).

Existing studies have identified some factors associated with purchasing products from overseas websites. These factors are: affordable prices, better products with a greater range, and high quality services (Forrest and Sullivan 2012). Also, overseas websites provide a free or lower delivery cost (Forrest and Sullivan 2012; PwC 2012). Moreover, a strong Australian dollar (Moodie 2012), and GST exemption for under- AU$1000 overseas purchases on-line (Irvine et al. 2011; Keating 2014) influence this.

Our study differs from these studies, in that they focus on consumers' buying decision associated with different factors. We focused on improving managerial strategic-decision support requirement in the small business B2C e-commerce sector, so owner/managers could have better options for attracting consumers. It was recognised that owner/managers are the most important resource within this type of organisation and their commitment or decisions are the most influential in shaping the performance of an organisation (Hansen and Hamilton 2011; Mazzarol et al. 2009). We did not focus on matters of truth and consumers' motivation or buying power factors. Previous studies did not focus on the small business sector in Australia, where over two million small businesses exist. They comprise 95.9 percent of all businesses, according to the Office of the Australian Small Business Commissioner (ASBC 2013). Small businesses are major contributors to national economies around the world. Therefore, the need for appropriate support for making strategic decisions related to web-features selection is an important aspect in order to address the vital issue – of products being purchased from overseas websites instead of local small business B2C websites. This represents a clear need for a potential solution in this regard. Thus, the main research questions are:

1. What are the significant factors that influence consumers to purchase products from overseas websites rather than local small business B2C websites?
2. What would be a solution that may help small business owner/managers in making strategic decisions about selecting relevant, competitive features on their websites to attract consumers to make purchases from them?

DSS research can provide potential solutions to organizational problems. DSS assists in improving managerial strategic decision-making (Arnott and Pervan 2012), and can help decision makers understand the issues underlying a system (Wienclaw 2008). It also assists decision makers to make better decisions (Pick 2008; Wienclaw 2008). A "system is a collection of interrelated components that work together towards a collective goal" (Chaffey and Wood 2005, p.26). Components of a system can be people, technology, and the environment involved with the business process, and provides an effective outcome (Atiq et al. 2011). When managers make decisions in an environment where markets change rapidly, and consumer demands are increasingly high; thus it is difficult to forecast the online environment and attract customers (Power et al. 2007). Appropriate decisions and actions must be taken by decision makers in this environment, as interactions increase among business environmental factors (Hall 2008). Therefore, the employment of DSS becomes not only desirable but is considered essential for business decision-makers (Alalwan 2013).

## 3. RESEARCH DESIGN: THEORETICAL FRAMEWORK

Typically, e-commerce systems have certain components with relevant website features including interactive marketing, ordering, payment and customer support processes, with a real-time catalogue (Baltzan and Phillips 2009). However, many small business website features in the B2C sector are very different from medium and large-sized business sites. The website represents the world marketplace for visitors searching for products and services. Accessible websites can give businesses many benefits, including increased visibility, extensive business areas, and enhanced customer service, which create customer satisfaction. When creating interactive websites, many businesses offer e-commerce websites that provide virtual web fonts and multimedia catalogues, interactive order processing, secure payment systems and online customer support (Marakas and O'Brien 2014). Website features in the virtual environment can be categorised on business information (I), communication (C), transaction (T), and distribution (D) provision (Angehrn 1997).

Therefore, in the virtual web environment, business information should contain useful features on their websites regarding visibility and access to product and service information. These include information on marketing, advertising and pricing. Good business communication space contains useful information for developing, monitoring and influencing business-related communication. This includes relationship building, lobbying, email, call centre information, feedback and login, through the creation of virtual communities (e.g., potential and existing customers, partners, and competitors) on their websites. Furthermore, virtual transaction web features contain useful information regarding initiation and execution of transactions, including orders and payments. Virtual distribution web features contain useful information regarding web-based distribution of products and services comprising digital goods and content, software and online physical-product tracking information (Angehrn 1997). These website features are basic requirements for interacting with consumers in the B2C e-commerce environment.

However, more than half the small businesses in Australia have used websites only for promoting their businesses (Fisher et al. 2007; Sensis 2014). Many small business owners view their websites as simply an advertising medium (Fisher et al. 2007). Domestic online shopping options have been limited, due to slow investment in Australian B2C ecommerce (Irvine et al. 2011). A recent study has also found that although ninety-five per cent of SMEs in Australia have an online presence, only nineteen per cent have a digital strategy, including Internet, website, social and mobile components (Snesis 2014). Moreover, in a current report provided by the Australian Bureau of Statistics (ABS 2015), only seven per cent of small businesses have automated links between systems used to receive orders, and other business systems. These pieces of evidence demonstrate lack of use of relevant features on small businesses websites. Features such as ordering, payment and customer-support related options are significant factors both for attracting and interacting with consumers on a business website. Therefore, a proposition has been formed:

***Proposition I:*** *There is a link between relevant features that should be available for small business B2C websites, and the factors that influence consumer purchasing from overseas websites, such as competitor websites.*

Although a few studies in the context of Australian B2C ecommerce have involved research on consumer buying decisions, these studies have not focused on factors related to website features. Relevant studies focus on demographics (Chang and Samuel 2004), convenience, price, time-saving, ease of comparison of products and prices (Irvine et al. 2011; Michael 2006), and enjoyment (Islam and Miah 2012). The goal of these studies was to determine the relationship between consumers' motivational factors when online shopping, and attitudes to shopping. However, our study explored the relationships between the factors that influence online consumer purchasing from overseas websites and small business owner/manager strategic decisions about the selection of relevant features on their B2C websites.

Previous studies also found that only a minority of small businesses use strategic planning (Blackburn et al. 2013; Mazzarol et al. 2009). Small businesses are constrained by a lack of resources (time, finance, and expertise) and lack of strategic decision- making, when focusing on planning effective use of IT. Some small business owner/managers have a business strategy, but it is not often related to their web presence and website matters (Fisher et al. 2007). This represents a lack of strategy, particularly related to assistance in setting up features on their websites. Hence, a proposition has been formed:

***Proposition II:*** *There is a link between relevant features that influence consumer purchases from overseas websites, and small business owner/managers' strategic decision-making about the selection of which are the relevant features to be available on their B2C websites.*

However, such decision-making is difficult for small business managers without the use of a DSS tool. In a rapidly changing environment, decision makers can be influenced in several ways. When managers make decisions in an environment where markets change rapidly, and consumer demands are increasingly high, it is sometimes difficult to forecast the environment (Power et al. 2007), and retention of consumers. Appropriate decisions and actions are required by decision makers in changing environments (Hall 2008). The need for quality and real-time decision-making is paramount as the business environment grows more complex and competitive. Therefore, the employment of DSS is essential for business decision-makers (Alalwan 2013). Hence, a proposition has been proposed:

***Proposition III:*** *There is a great necessity to design and develop a DSS solution framework that will support small business owner/managers in making strategic decisions to select relevant features on their B2C e-commerce websites.*

To address the research questions raised, we developed the above propositions. The next section describes the research methodology and the different phases of data collection process, intended to design and develop a practical DSS solution framework by employing a design science research method. Building from Arnott and Pervan's (2012) DSS design, this study adopted design science (DS) principles from the DSS artefact defined by Miah et al. (2014). In DS principle, Hevner et al.'s (2004) seven guidelines, as shown in Table 1, provide supportive insights for defining a problem space, outlining and implementing a design and evaluating a design artefact for the proper communication of research in a more user-centered way. This design science principle has promising options that provide further clarity for constructing DSS artefacts in business or natural settings.

## 4. RESEARCH METHODOLOGY

To achieve our research aims, we employed a constructivist research philosophy. Constructivism research provides practical guidelines for understanding a context with multiple perspectives and diversities, and generates theory (Creswell 2013, p.8). In theory building, "ontology is a philosophical approach to theory building based on investigating the universal and necessary characteristics of all existence" (Lancaster 2005, p. 35). In the same concept, "epistemology is a philosophical approach to theory building which investigates the nature, grounds, limits and validity of human knowledge" (Lancaster 2005, p. 35).

The constructivist research approach often "combines with interpretivism" and is typically seen as an approach to qualitative research (Creswell 2013, p. 8). In this approach, individual researchers develop subjective meanings from their experiences that direct toward a certain object or thing. The goal of the researcher is to rely on the participants' views of the situation being studied. The questions are therefore, become increasingly broad, general and open-ended. Thus, participants can construct the meaning of a situation. Therefore, we employed a qualitative research approach for our study.

To answer our research questions, that is, to achieve the objectives of the research (evaluating propositions, as it is inductive research); the research methodology consisted of three phases for the data collection processes.

### 4.1　Phase one: literature review

This phase involved critical reviews of relevant literature in the field of e-commerce, small business B2C e-commerce, and DSS in the IS discipline. Two main research questions and three propositions emerged from this phase; investigating the research problem and providing a potential solution. The phase is completed in our study.

### 4.2　Phase two: case studies and web content analysis

For the improvement of web features, this phase involved investigating internal and external business environmental factors of a small business in the B2C sector that relates to owner/managers' strategic decisions and consumers purchasing from overseas websites. A case study method will be used, with multi-cases for the data collection to identify the internal business environmental factors (this data collection will start after the ethics approval). We already employed web content analysis of the data collection to identify the external business environmental factors.

A case study is an empirical inquiry that investigates a contemporary phenomenon within its real-life context (Yin 2009). A case-study design should be considered when the focus of the study is to answer "what," "why," "how" and "when" questions (Baxter and Jack 2008). Our study will be used an inductive case-study method that is an iterative process, where data is will be analysed and compared with the existing theory to develop the new theory (Myres 2009). Here, the theoretical framework has been built for further theory development through inductive case-study analysis. We will examine twenty small businesses in the B2C sector in Australia that have web presence, using a purposive sampling. We will use a qualitative data collection approach, because case studies are the most common qualitative method used in business and IS research. Case studies are based on contemporary issues, and the cases document one or more organisations, to deal with issues that are important to other organisations (Myers 2013). Interviews are a useful qualitative data collection technique. This technique is used for a variety of purposes including need assessment, issue identification, and strategic planning (Guion et al. 2011; Mintzberg et al. 1976). The data will be collected from small businesses in the B2C sector throughout Australia via email interviews with owner/managers, by employing semi-structured interview questions. This data collection has not yet to commence.

To identify important external business environmental factors, we examined twenty small businesses' website features, comparing five of their competitors' websites that are selling to Australia, through Web content analysis. This is also a purposive sampling.

### 4.3　Phase three: solution design and evaluation

This phase involves the development and evaluation of a DSS solution framework. A prototype method will be used in conjunction for DSS design, to target owners/managers. A prototype is a smaller-scale representation or working model of the user's requirements for a proposed design (Haag and Cummings 2009). For an ensemble artifact design (Miah and Gammack, 2014) we adopted the design science research methodology (DSRM) by employing Hevner et al.'s (2004) DSR guidelines, consisting of seven phases of research design (see Table 1), where design science is inherently a problem-solving process (Hevner et al. 2004). The DSR guidelines are helpful to determine the information needed on relevant website features in small business B2C websites, and user engagements, such as owner/manager participation in the development process of the DSS solution framework.

| Guidelines of design research | Relevance to our artefact design research |
|---|---|
| Guideline 1: Design an Artefact | An innovative artefact with software solution prototype will be developed, with user involvement and specified as a reproducible DSS solution framework. |
| Guideline 2: Problem Relevance | A real business problem domain was defined (see Section 2) that produced two research questions. A theoretical framework (see Section 3) was developed to answer two research questions. The problem will found through a Web content analysis and case studies. For potential solutions, this study will be used Miah et al.'s (2014) user-centric DSS design model, through owner/managers' participation in the DSS design process. |
| Guideline 3: Design Evaluation | A descriptive evaluation method will be employed for prototype testing of the final DSS solution framework, through the involvement of the owner/managers from B2C small businesses in Australia who have online presence. |
| Guideline 4: Research Contributions | Academic contributions are: designed a theoretical framework and research methods (Web content analysis and case studies) that can assist in identifying real business problems. For a solution, a practical contribution of this study will be development of a DSS solution framework by employing a prototype. (See Figure 1, a proposed framework). |
| Guideline 5: Research Rigour | Research rigour was derived from the knowledge base through development of a theoretical framework. The theories partially were tested by employing web content analysis. However, these have not tested yet through multiple case studies. The results will be used for development of a DSS solution framework through the prototype method for evaluating the framework. |
| Guideline 6: Design as a search process | The method of artefact is closely aligned with owner/manager inputs. A DSS solution framework will be developed based on results from web content analysis and case studies, with the scanning of internal and external business environmental factors. |
| Guideline 7: Communication of research | This will be achieved through system demonstrations and evaluation by small business owner/managers. |

*Table 1. Seven principles adapted for conducting our DS research (Hevner et al. 2004, p. 83)*

## 5. DATA ANALYSIS AND RESULTS

This document is a research-in-progress paper. This study involves the different phases of the data collection process, their analysis and results. The following sections describe the progress of our study regarding data collection, analysis and results.

### 5.1 Data analysis in phase two: case study with small business owner/managers

This data collection has not commenced yet, so we do not have any findings to discuss. However, the findings data will be analysed and compared with the existing theory and a new theory developed, and this is the inductive data analysis method in qualitative research (Myres 2009, p. 75).

### 5.2 In phase two: web content analysis and results

| Overall small business website feature adoption levels (31.4%) | | | Virtual Space | Overall overseas website feature (competitors) adoption levels (88%). | | |
|---|---|---|---|---|---|---|
| Added value (23%) | Business information (35%) | Contact details (100%) | I | Contact details (100%) | Business information (100%) | Added value (93%) |
| Social networking (35%) | Reviews (5%) | Contact the business (31%) | C | Contact the business (60%) | Reviews (40%) | Social networking (100%) |
| Payment (17%) | Online database (28%) | | T | Online database (93%) | | Payment (90%) |
| E-service (20%) | Online distribution (20%) | | D | Online distribution (100%) | | E-service (100%) |

*Table 2. The summary of web content analysis results*

We have conducted a web content analysis on twenty small business websites within the B2C sector in Australia and five overseas websites who sell to Australia to test Proposition I. This is one of the investigation methods to identify the research problem. The results discussed in this section were based on the available features of small business websites in the B2C sector, in comparison with overseas websites who sell to Australia. The summary of content analysis and the visual display of this result has presented using the Angehrn's (1997) ICDT Model (see Table 2).

We categorised the website features that are necessary to attract potential consumers within the B2C online environment. The website features have been categorised based on the business information (I), communication (C), transaction (T) and distribution (D) processes in the virtual organisation context (Angehrn 1997). In Table 2, left side displays the adoption level of website features within small businesses. The right side displays the adoption level features in overseas websites who sell to Australia. The cross-case analysis technique was used to discuss the web content analysis (Yin 2012). Therefore, we analysed the results based on the features available on small business websites and compared these features with overseas website features through qualitative cross-case analysis. Our study found that the adoption level of website features within the small business B2C sector in Australia is significantly lower (31%) than the adoption level of overseas website features (88%). However, all small businesses have contact details, the same as overseas websites who sell to Australia. This partial result suggests that there is a requirement for a DSS solution framework. Small business owner/managers could use this framework to select relevant features on their websites to interact with potential consumers. The next section proposes and outlines a conceptual DSS solution framework.

## 6. PROPOSED DSS SOLUTION FRAMEWORK

We propose a conceptual DSS solution framework for a potential solution to the research problem raised. This DSS solution framework is based on Mintzberg et al.'s (1976) model for small business owner/managers' strategic decision-making within the B2C e-commerce environment. We extended this model by adapting Miah et al.'s (2014) user-centred DSS model, where the user-centred DSS meets the decision-makers' contextual needs in their businesses (Miah et al. 2014). Mintzberg et al.'s (1976) model consists of three phases: identification, development and selection phases (see Figure 1). Throughout these decision process phases, owner/managers can use methods to identify problems and opportunities, and find alternative paths for potential solutions to product being purchased from overseas websites. For a potential solution, Miah et al.'s (2014) user-centred DSS model will be used. These phases are described as follows.

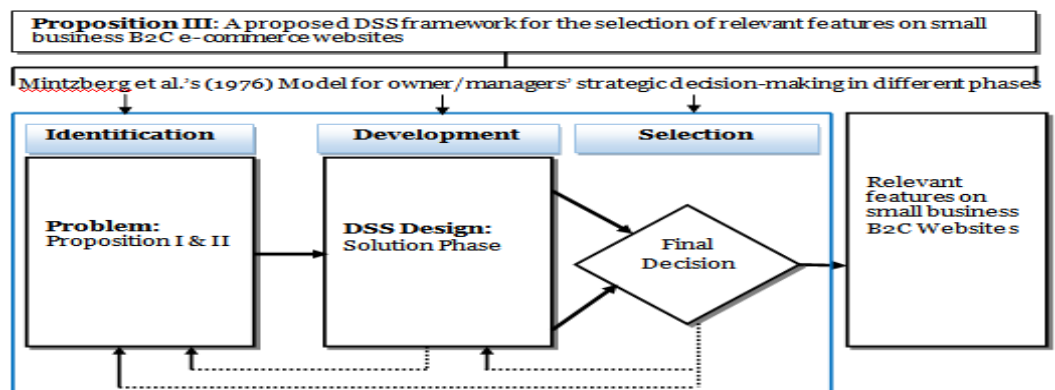

*Figure 1. A proposed DSS solution framework*

### 6.1 Identification phase

This phase comprises two routines: "recognition" and "diagnosis". The "recognition" routine initiates the decision process by recognising problems and opportunities. Hence, owner/managers can recognise problems through this routine, such as many Australian consumers purchasing products from overseas websites rather than small Australian business B2C websites, due to many factors as discussed. In the "diagnosis" routine, further information is required to define and clarify the previously recognised problem or opportunity. Therefore, information is required for owner/managers in strategic decision-making.

In the "diagnosis" routine, a strategic decision focuses on and analyses both internal and external business environmental factors (Mintzberg et al. 1976). Typically, internal environmental analysis involves identifying strategic factors crucial to the success of an organisation, determining the importance of each of these factors, determining the strengths and weaknesses of the organisation in each of these factors, and finally, preparing a strategic advantage profile for the organisation and comparing it with profiles of successful competitors in the industry (Srivastava and Verma 2012). Examples of internal factors are: skills and knowledge, technology use, and management strategies (Pedersen and Sudzina 2012), such as organisation's objectives, strengths and problems (Thierauf 1988, p. 236). The external business factors are: customer demands, general economic condition, regulations, new technology and competition (Pedersen and Sudzina 2012; Thierauf 1988). Therefore, this study considers small-business external factor analyses, and involves an examination of how competitors are using their websites. In this context, competitor website features (the overseas website features) related to consumer online purchase factors are considered to be external business environmental factors. Proposition I was developed to investigate these external business environmental factors. The internal factors are small business owner/managers' knowledge, skills and strategies for the use of advanced technology. Also, their strategic decisions are internal factors in selecting relevant website features on their websites. Therefore, Proposition II has been proposed for investigating these internal business environmental factors through the case study method, with multiple cases of small business owner/managers in the B2C sector in Australia.

## 6.2 Development phase

This phase involves a set of activities that generates one or more solutions. This phase has two routines. First is the search routine that aims at finding ready-made solutions. Second is the design routine that aims to develop new solutions or modify ready-made ones. In this phase, owner/managers can formulate or design a decisions model, set the criterion for the choice, and search for alternatives. Therefore, Miah et al.'s (2014) user-centred DSS model will be be used for developing a practical DSS solution framework with small business owner/managers' participation.

## 6.3 Selection phase

Mintzberg et al. (1976) suggested the "Selection Phase" (see Figure 1) is typically a multi-stage iterative process of decision making. Three routines emerged from this phase: screen, evaluation-choice and authorisation. The selection phase starts with a "screening routine", which is activated to eliminate any impractical alternatives. Next, the best alternative is selected, through a process of analysis, in the "evaluation-choice routine". Finally, the decision goes through the "authorisation routine". This routine involves an authorised decision-maker for making the strategic decision that relates to the selection of competitive website features. Competitive features for small business websites are categorised based on the business information (I), communication (C), transaction (T) and distribution (D) processes in the virtual organisation context by Angehrn (1997), as discussed early in Section 3.

# 7. DISCUSSION AND CONCLUSIONS

In this paper, we described a research-in-progress study that aims to outline requirements of an innovative DSS solution framework that informs from case study findings with small business owner/managers and a content analysis of website features. This framework shows promise in assisting small business owner/managers in making strategic decisions concerning the selection of competitive features on their B2C websites, to reach potential consumers. The DSS will be a personal DSS type, a small-scale system developed for one manager or a small number of independent users in order to support a decision-making task (Arnott and Pervan 2008). Miah, Kerr & Gammack (2009) describe three directions of domain specific DSS such as advisory systems; diagnostic systems and planning & management support systems. We aimed to develop the DSS that may focus on meeting the strategic demand through supporting appropriate design information.

We mainly outlined the theoretical basis of the solution in this paper. The relationship between owner/managers' strategic decision-making about the selection of competitive features on their websites and relevant features that influence consumer purchases from overseas websites has been outlined by designing a theoretical basis. This theoretical basis formed proposition I and proposition II, through critical reviews of the literature to investigate the research problem. The relationship will be determined practically through evaluating these two propositions by employing case studies and web content analysis. We identified the relationship between the factors but yet to be employing case studies for achieving more empirical justifications.

Along with our plan to conduct multi-case studies in order to find internal factors in the business environment, our study also conducted the web content analysis, which we reported to understand the external factors related to web features (both local and overseas websites, for comparison). Our study found that the adoption level of the features of small business website is significantly lower (31%) than the features adoption of level of overseas websites (88%) selling to Australia. However, all small businesses have contact details on their websites (see Table 2). Using both internal and external factors, we outlined the DSS solution framework in this paper. The initial results from the web content analysis clearly suggest a need for a new DSS solution design.